\documentclass[prl,twocolumn,floatfix,superscriptaddress]{revtex4-1}
\usepackage{dcolumn,amsmath}
\usepackage{graphicx}
\usepackage{bm}
\usepackage{hyperref}

\newcommand{\ecm}{\ensuremath{e {\cdotp} {\rm cm}}}
\newcommand{\eEDM}{{\em e}EDM}
 \newcommand{\Eeff}{\mathcal{E}_\mathrm{eff}}
\newcommand{\B}{\mathcal{B}} %magnetic field
\newcommand{\E}{\mathcal{E}} %electric field
\newcommand{\Bvec}{\vec{\mathcal{B}}} %magnetic field
\newcommand{\Evec}{\vec{\mathcal{E}}} %electric field
\newcommand{\de}{d_\mathrm{e}}

\begin{document}
%  \title{Zeeman interaction in ThO $H^3\Delta_1$}
   \title{AC Stark effect in ThO $H^3\Delta_1$ for the electron EDM search}
\author{A.N.\ Petrov}\email{alexsandernp@gmail.com}
\affiliation
{National Research Centre ``Kurchatov Institute'' B.P. Konstantinov Petersburg
Nuclear Physics Institute, Gatchina, Leningrad district 188300, Russia}
\affiliation{Division of Quantum Mechanics, St.Petersburg State University, 198504, Russia}
%\date{}
\begin{abstract}
A method and code for calculations of diatomic molecules in the external variable electromagnetic field have been developed.
Code applied for calculation of systematics in the electron's electric dipole moment search experiment on ThO $H^3\Delta_1$ state related to geometric phases,
including dependence on $\Omega$-doublet, rotational level, and external static electric field.
It is found that systematics decrease cubically with respect to the frequency of the rotating transverse component of the electric field.
Calculation confirms that experiment on ThO $H^3\Delta_1$ state is very robust against systematic errors related to geometric phases.
\end{abstract}

\maketitle

%========================================================================
%\section{EDM measurements with $\Omega$-doublets}

The experimental measurement of a non-zero electron electric dipole moment (\eEDM, $\de$) would be a clear signature of physics beyond the Standard model  \cite{Khriplovich1997,Pospelov2005,Titov:06amin, Feng:2013}.
The current limit for \eEDM, $|\de|<9\times 10^{-29}$ \ecm\ (90\% confidence), was set with a buffer-gas cooled molecular beam\cite{Baron2013,Hutzler2012,Patterson2007} of thorium monoxide (ThO) molecules in the metastable electronic $H^3\Delta_1$ state.
It was shown that due to existence of closely-spaced levels of opposite parity of $\Omega$-doublet the experiment on ThO is very robust against a number of systematic effects related to magnetic fields\cite{DeMille2001, Petrov:14} or geometric phases\cite{Vutha2009}. However, the upper and lower $\Omega$-doublet states have slightly different properties and systematic effects related to magnetic field imperfections and geometric phases can still manifest themselves as a false {\eEDM}. 
 The dependence of $g$-factors of the ThO $H^3\Delta_1$ state on $\Omega$-doublets and external electric field was considered in Ref. \cite{Petrov:14}. The aim of the present
work is to consider geometric phase shifts.
%
%\section{Theory}
%

Following the computational scheme of \cite{Petrov:11, Petrov:14}, the energies of the rotational levels in the $H^3\Delta_1$ electronic state of the $^{232}$Th$^{16}$O molecule in external {\it static} electric $\Evec = \E\hat{z}$ and magnetic $\Bvec = \B\hat{z}$ fields are obtained by numerical diagonalization of the molecular Hamiltonian (${\rm \bf \hat{H}}_{\rm mol}$)  over the basis set of the electronic-rotational wavefunctions $\Psi_{\Omega}\theta^{J}_{M,\Omega}(\alpha,\beta)$. Here $\Psi_{\Omega}$ is the electronic wavefunction, $\theta^{J}_{M,\Omega}(\alpha,\beta)=\sqrt{(2J+1)/{4\pi}}D^{J}_{M,\Omega}(\alpha,\beta,\gamma=0)$ is the rotational wavefunction, $\alpha,\beta,\gamma$ are Euler angles, and $M$ $(\Omega)$ is the projection of the molecule angular momentum ${\bf J}$ on the lab $\hat{z}$ (internuclear $\hat{n}$) axis.
Detailed feature of the Hamiltonian is described in \cite{Petrov:14}. 
  In the paper the $M = \pm 1$ states which represent interest for \eEDM\ search experiment are considered. For electric field $\E = 20 - 200~{\rm V/cm} $, used in the experiment, lower rotational levels with $M \ne 0$ can be labeled by $|J, M,\Omega>$ quantum numbers. States $|J,M {=}1,\Omega{=}1>$, $|J, M{=}-1,\Omega{=}-1>$ correspond to the upper and $|J, M{=}-1,\Omega{=}1>$, $|J, M{=}1,\Omega{=}-1>$ to the lower $\Omega$-doublet levels.
External magnetic field removes the degeneracy between $\Omega$-doublet components: $\Delta E^u = E(|J,M{=}1,\Omega{=}1>) - E(|J, M{=}-1,\Omega{=}-1>)$, $\Delta E^l = E(|J, M{=}1,\Omega{=}-1>) - E(|J, M{=}-1,\Omega{=}1>)$.
The relevant energy levels can be seen in Figure 2 of Ref. \cite{Skripnikov:13c} or Figure 3 of Ref. \cite{Vutha:2010}.
Provided g-factors for upper and lower $\Omega$-doublet levels are close enough $\Delta E^u$ and $\Delta E^l$ remain
equal unless both parity and time reversal symmetries are violated. The difference in splitting gives the value for \eEDM\ $ d_e = \frac{|\Delta E^l -\Delta E^u |}{4\Eeff}$, here $\Eeff = 81.5 {\rm GV/cm}$ \cite{Skripnikov:13c, Skripnikov:14b} is the effective 
internal electric field. However, there are systematic effects which can give additional energy shifts $\delta\Delta E^l$ and  $\delta\Delta E^u$ for $\Delta E^l$ and $\Delta E^u$ which manifest as a false {\eEDM}.
This leads to a systematic error $ \delta d_e(sys) = \frac{\delta \Delta E^l - \delta \Delta E^u }{4\Eeff}$.
It is also useful to consider systematic effects $ \tilde{\delta} d_e(sys) = \frac{\delta \Delta E^{l(u)} }{4\Eeff}$ related to one of the $\Omega$-doublet component .
One of the effect is the interaction with transverse component of the electric field $\vec{\E}(t) = \E_{\perp}(\hat{x}cos(\omega_{\perp}t) + \hat{y}sin(\omega_{\perp}t))$ which
appears due to a spatial inhomogeneties in the applied electric field \cite{Vutha:2010}.
Let us consider this effect.

The corresponding part of the Hamiltonian is
\begin{equation}
{\rm {\bf\hat{H}}_{tf}} = -\vec{d}\cdot\vec{\E}(t) = -\E_{\perp}/2(d_{+}e^{-i\omega_{\perp}t} +  d_{-}e^{i\omega_{\perp}t}),
\end{equation}
where $d_{\pm} = d_{x} \pm id_{y}$.  It is more convenient to describe the interaction of the molecule with the quantized
electromagnetic fields. The corresponding Hamiltonian is
\begin{equation}
{\rm {\bf\hat{H}}_{int}} = \hbar\omega_{\perp}a^+a - \sqrt{\frac{2\pi\hbar\omega_{\perp}}{V}}(d_{+}a^+ + d_{-}a),
\label{Hquant}
\end{equation}
where $a^+$ and $a$ are creation and annihilation operators, $V$ is a volume of the system.
To work with Hamiltonian (\ref{Hquant}) one need to add the quantum number $|{n}>$, where $n=\frac{V\E_\perp^2}{8\hbar\omega_{\perp}}$ is number of photons.
The approach is similar to the formalism outlined in Ref. \cite{Vutha2009}.
For this paper we consider the case $\E_{\perp} << \E$, such that the additional energy shifts
can be calculated in the framework of the second order perturbation theory.
%\begin{eqnarray}
\begin{align}
\nonumber
\delta\Delta E^u  =  \E_{\perp}^2/4 \times \\
\nonumber
 ( \sum_{J',\Omega'} \frac {|<J, M{=}1,\Omega{=}1|d_{+}|J',M{=}0,\Omega'>|^2}{E(J, M{=}1,\Omega{=}1)-E(J',M{=}0,\Omega')+\hbar \omega_{\perp}}  \\
\nonumber
+\sum_{J',\Omega'} \frac {|<J, M{=}1,\Omega{=}1|d_{-}|J',M{=}2,\Omega'>|^2}{E(J, M{=}1,\Omega{=}1)-E(J',M{=}2,\Omega')-\hbar \omega_{\perp}} \\
\nonumber
-\sum_{J',\Omega'} \frac {|<J, M{=}-1,\Omega{=}-1|d_{+}|J',M{=}-2,\Omega'>|^2}{E(J, M{=}-1,\Omega{=}-1)-E(J',M{=}-2,\Omega')+\hbar \omega_{\perp}} - \\
-\sum_{J',\Omega'} \frac {|<J, M{=}-1,\Omega{=}-1|d_{-}|J',M{=}0,\Omega'>|^2}{E(J, M{=}-1,\Omega{=}-1)-E(J',M{=}0,\Omega')-\hbar \omega_{\perp}} ),
\label{pt2u}
\end{align}
%\end{eqnarray}

\begin{align}
\nonumber
\delta\Delta E^l  =  \E_{\perp}^2/4 \times \\
\nonumber
 ( \sum_{J',\Omega'} \frac {|<J, M{=}1,\Omega{=}-1|d_{+}|J',M=0,\Omega'>|^2}{E(J, M{=}1,\Omega{=}-1)-E(J',M{=}0,\Omega')+\hbar \omega_{\perp}} \\
\nonumber
+\sum_{J',\Omega'} \frac {|<J, M{=}1,\Omega{=}-1|d_{-}|J',M{=}2,\Omega'>|^2}{E(J, M{=}1,\Omega{=}-1)-E(J',M{=}2,\Omega')-\hbar \omega_{\perp}} \\
\nonumber
-\sum_{J',\Omega'} \frac {|<J, M{=}-1,\Omega{=}1|d_{+}|J',M{=}-2,\Omega'>|^2}{E(J, M{=}-1,\Omega{=}1)-E(J',M{=}-2,\Omega')+\hbar \omega_{\perp}} \\
-\sum_{J',\Omega'} \frac {|<J, M{=}-1,\Omega{=}1|d_{-}|J',M{=}0,\Omega'>|^2}{E(J, M{=}-1,\Omega{=}1)-E(J',M{=}0,\Omega')-\hbar \omega_{\perp}} ).
\label{pt2l}
\end{align}
Major contribution to $ \delta\Delta E^{u(l)}$ comes from coupling of states with the same $J$. The most simple is the picture for $J{=}1$ state.
 $|J{=}1,M{=}1,\Omega,n>$ interact with $|J{=}1,M{=}0,\Omega,n{+}1>$ and $|J=1,M{=}-1,\Omega,n>$ with $|J,M{=}0,\Omega,n{-}1>$.
 Note, that ${\rm \bf \hat{H}}_{\rm mol}$ can only couple the states with the same $n$, whereas ${\rm {\bf\hat{H}}_{int}}$ couples the states with $\Delta n = \pm 1$.
Energies of states $|J,M{=}0,\Omega,n{+}1>$ and $|J,M{=}0,\Omega,n{-}1>$ are different by $2\hbar \omega_{\perp}$, 
and states $|J{=}1,M{=}1,\Omega,n>$ and $|J{=}1,M{=}-1,-\Omega,n>$ by $2\mu\B$. This leads to different energy denominators in eqs. (\ref{pt2u},\ref{pt2l}) and results in different energy shift
for $|J{=}1,M{=}1,\Omega,n>$ and $|J{=}1,M{=}-1,-\Omega,n>$.
%what gives main contribution to $ \delta\Delta E^{u(l)}$.
However, for $J{=}1$ level, it was shown in Ref. \cite{Vutha:2010} that (considering the interaction with $|J{=}1,M{=}0,\Omega>$ states only) provided tensor Stark
($\Delta E_{ST}$ = E(J,M{=} $\pm$1,$\Omega$) - E(J, M{=}0,$\Omega$))
 and Zeeman splitting are the same
for upper  and lower  component of the $\Omega$-doublet the AC Stark shifts $ \delta\Delta E^{u}$ and $\delta\Delta E^{l}$ will also be equal. This allows one to reject systematic errors due to geometric phases
by performing measurements in both $\Omega-$doublet states.
However the tensor Stark splittings do not coincide  exactly. Also including interaction with other states lift the degeneracy. It is particularly important to include
the interaction with the neighbor rotational levels. The latter interaction increases  value for $\delta d_e(sys)$
on several orders of magnitude whereas $ \tilde{\delta} d_e(sys)$ is almost unaffected by this interaction.
See also influence of perturbation by $J=2$ level on $J=1$ g-factors in Refs. \cite{Bickman2009,Petrov:11,Lee:13a,Petrov:14}.

Tables \ref{dde} and \ref{dde2} list the calculated $ \tilde{\delta} d_e(sys)$ and $\delta d_e(sys)$ as a functions of $\omega_{\perp}$ and $\E$ for $J=1$ and $J=2$, correspondingly.
Though for smaller $\E$ the $\E_{\perp}$ value will be smaller as well, for the calculation I take the same $\E_{\perp}=10 {\rm mV/cm}$
given in Ref. \cite{Vutha:2010} for all $\E$. Using the fact that $ \tilde{\delta} d_e(sys)$ and  $\delta d_e(sys)$ are quadratic functions of $\E_{\perp}$ the results can be easily recalculated for any $\E_{\perp}$.
For static magnetic field the value $\B=40 {\rm mG}$ used in the experiment \cite{Baron2013} is used.
One can see that $\delta d_e(sys)$ is two orders of magnitude larger for $J{=}2$ than for $J{=}1$ though much smaller than the current limit on $\de$.

Calculation for $\omega_{\perp}/2\pi$ less than 250 kHz is not performed due to the limited computational accuracy.
For smaller $\omega_{\perp}$ one can expect further decreasing of $ \tilde{\delta} d_e(sys)$ and $\delta d_e(sys)$.
Each term in Eqs. (\ref{pt2u},\ref{pt2l}) has form $\frac{b_{u(l)}^2}{a_{u(l)}+\hbar \omega_{\perp}} - \frac{b_{u(l)}^2}{a_{u(l)}- \hbar \omega_{\perp}}$.
Retaining terms up to the third order in $\omega_{\perp}$ we have
\begin{equation}
\delta\Delta E^{u(l)} \approx  -2\frac{B_{u(l)}^2}{A_{u(l)}} \left( \frac{\hbar \omega_{\perp}}{A_{u(l)}}  +  \frac{ \hbar^3 \omega_{\perp}^3}{A_{u(l)}^3} \right).
\label{assimp}
\end{equation} 
Formulae for $B_{u(l)}$ and $A_{u(l)}$ for $J{=}1$ are given below.
Eq. (\ref{assimp}) explains the fact that $\delta\Delta E^{u(l)}$ decreases linerly with small $\omega_{\perp}$ listed in Tables \ref{dde} and \ref{dde2}.
Similarly to $ \delta\Delta E^{u(l)}$ the major contribution to difference $ \delta\Delta E^{l} - \delta\Delta E^{u}$ comes from coupling of states with the same $J$
(terms with $J'=J$ in Eqs. (\ref{pt2u},\ref{pt2l})).
However, important role plays the perturbation by the closest rotational levels which makes matrix elements and denominators for upper and lower components of $\Omega$-doublet slightly different.
Let us consider this effect for the simplest case the $J{=}1$ level. Without perturbation by the $J=2$ level the parameters
\begin{align}
\label{stark}
\nonumber
A = \Delta E_{ST} = -\E <J{=}1,{M{=}1,\Omega}|d_z| J{=}1,{M=1}, \Omega>  \\
= -\E d M\Omega/J(J{+}1)
\end{align}
and
\begin{align}
\nonumber
B =   -\E_{\perp}/2 <J{=}1,{M{=}1,\Omega}|d_+| J{=}1,{M'{=}0}, \Omega> \\
=  -\frac {\E_{\perp} d \Omega} {2} \frac {\sqrt{(J{-}M{+}1)(J{+}M)}} {J(J{+}1)}
\end{align}
up to the sign are the same for upper and lower $\Omega$-doublet levels. $\Delta E_{ST}$
is positive for upper and negative for lower levels. Note that dipole moment $d < 0$.
Eqs. (\ref{assimp},\ref{stark}) explain the fact that $\delta\Delta E^{u(l)}$ decreases quadratically with $\E$.

\begin{table}
\caption{ 
The $ \tilde{\delta} d_e(sys)$ (in units 10$^{-29}$\ecm\ ) and $ \delta d_e(sys)$ (in units 10$^{-34}$\ecm\ ) calculated for the $J=1$ $H^3\Delta_1$ state in $^{232}$Th$^{16}$O.}
\begin{tabular}{ccccccc}
\hline
 $\E$  & \multicolumn{2}{c} {$\omega_{\perp}/ 2\pi=4$MHz} &  \multicolumn{2}{c} {$\omega_{\perp}/ 2\pi=1$MHz} & \multicolumn{2}{c} {$\omega_{\perp}/ 2\pi=250$kHz}\\

    (V/cm)    & $ \tilde{\delta} d_e(sys)$ & $ \delta d_e(sys)$ &   $ \tilde{\delta} d_e(sys)$ & $ \delta d_e(sys)$ &  $ \tilde{\delta} d_e(sys)$ & $ \delta d_e(sys)$  \\
\hline
   20.  &  -1299.  &  -2144.  &   -313.   & -31.      &   -78.     &  -0.48     \\ 
   30.  &   -565.  &   -608.  &   -139.   & -9.1      &   -35.     &  -0.14     \\
   40.  &   -315.  &   -253.  &    -78.   & -3.8      &   -19.     &  -0.059    \\      
   50.  &   -201.  &   -128.  &    -50.   & -1.9      &   -12.     &  -0.030    \\
   60.  &   -139.  &   -74.   &    -35.   & -1.1      &   -8.6     &  -0.018    \\
   70.  &   -102.  &   -47.   &    -25.   & -0.70     &   -6.3     &  -0.011    \\
   80.  &    -78.  &   -31.   &    -19.   & -0.47     &   -4.8     &  -0.0073   \\
   90.  &    -62.  &   -22.   &    -15.   & -0.34     &   -3.8     &  -0.0053   \\
  100.  &    -50.  &   -16.   &    -12.   & -0.24     &   -3.1     &  -0.0038   \\
  110.  &    -41.  &   -12.   &    -10.   & -0.18     &   -2.6     &  -0.0028   \\
  120.  &    -35.  &   -9.2   &    -8.6   & -0.14     &   -2.2     &  -0.0022   \\
  130.  &    -30.  &   -7.3   &    -7.3   & -0.11     &   -1.8     &  -0.0017   \\
  140.  &    -26.  &   -5.8   &    -6.3   & -0.088    &   -1.6     &  -0.0014   \\
  150.  &    -22.  &   -4.7   &    -5.5   & -0.071    &   -1.4     &  -0.0011   \\
  160.  &    -20.  &   -3.9   &    -4.9   & -0.059    &   -1.2     &  -0.00093  \\
  170.  &    -17.  &   -3.2   &    -4.3   & -0.050    &   -1.1     &  -0.00078  \\
  180.  &    -15.  &   -2.7   &    -3.8   & -0.043    &   -0.96    &  -0.00067  \\ 
  190.  &    -14.  &   -2.3   &    -3.4   & -0.034    &   -0.86    &  -0.00053  \\
  200.  &    -12.  &   -2.0   &    -3.1   & -0.030    &   -0.77    &  -0.00047  \\
\hline
\end{tabular}
\label{dde}
\end{table}

Perturbation by $J=2$ changes the parameters:
\begin{equation}
A_{u(l)} = \Delta E_{ST} + \delta^{1} \Delta E_{ST}^{u(l)} + \delta^{2} \Delta E_{ST}^{u(l)},
\end{equation}

\begin{equation}
 B_{u(l)} = B + \delta^{1}B_{u(l)} + \delta^{2}B_{u(l)}.
\end{equation}
$\delta^{1(2)} \Delta E_{ST}$ is the correction to $\Delta E_{ST}$ due to
shifting down $|J{=}1,M{=}\pm1,\Omega>$ ($|J{=}1,M{=}0,\Omega>$) levels
when interacting with $J{=}2$. $\delta^{1} \Delta E_{ST}^{u(l)}$
is negative. It decrease (increase absolute value) $\Delta E_{ST}$ for upper
(lower) $\Omega$-doublet levels. In turn $\delta^{2} \Delta E_{ST}^{u(l)}$
is positive. It increases (decreases absolute value) $\Delta E_{ST}$ for upper
(lower) $\Omega$-doublet levels.
$\delta^{1(2)}B$ is the correction to $B$ due to the perturbation of the wavefunction
$|J{=}1,M{=}\pm1,\Omega>$ ($|J{=}1,M{=}0,\Omega>$) by $|J{=}2,M{=}\pm1,\Omega>$ ($|J{=}2,M{=}0,\Omega>$) one.
It is shown in 
%(\ref{app}) 
APPENDIX
that for $J{=}1$ level the corrections $\delta^{1} \Delta E_{ST}^{u(l)}$, $\delta^{2} \Delta E_{ST}^{u(l)}$,
$\delta^{1}B_{u(l)}$, and $\delta^{2}B_{u(l)}$ are correlated in such a way that
\begin{equation}
\frac {B_{u(l)}^2} {A_{u(l)}^2}  = \frac {B^2} {A^2}.
\label{cancel}
\end{equation}
Eq. (\ref{cancel}) is correct up to the second order in small parameter $\Delta E_{ST}/ \Delta E_{rot}$, where
$\Delta E_{rot} = E(J{=}1) - E(J{=}2)$ is energy difference between the first and second rotational levels.
Due to Eq. (\ref{cancel}) the linear term in the difference $ \delta d_e(sys) = \frac{\delta \Delta E^l - \delta \Delta E^u }{4\Eeff}$
is canceled and $ \delta d_e(sys)$, in the leading order, is a cubic function of $\omega_{\perp}$ for $J{=}1$. This behavior can be seen from the data in
Table \ref{dde}. Dependence of the $ \delta d_e(sys)$ for $J{=}2$ level on  $\omega_{\perp}$ has also nearly the cubic character.

The calculations confirm that the experiment on ThO $H^3\Delta_1$ state is very robust against systematic errors related to geometric phases.
Developed code can be applied for calculation of molecules in an ion trap at presence of rotating field \cite{Leanhardt:2011,Cornell:13}.

\begin{table}
\caption{ 
The $ \tilde{\delta} d_e(sys)$ (in units 10$^{-29}$\ecm\ ) and $ \delta d_e(sys)$ (in units 10$^{-34}$\ecm\ ) calculated for the $J=2$ $H^3\Delta_1$ state in $^{232}$Th$^{16}$O.}
\begin{tabular}{ccccccc}
\hline
 $\E$  & \multicolumn{2}{c} {$\omega_{\perp}/ 2\pi=4$MHz} &  \multicolumn{2}{c} {$\omega_{\perp}/ 2\pi=1$MHz} & \multicolumn{2}{c} {$\omega_{\perp}/ 2\pi=250$kHz}\\

    (V/cm)   & $ \tilde{\delta} d_e(sys)$ & $ \delta d_e(sys)$ &   $ \tilde{\delta} d_e(sys)$ & $ \delta d_e(sys)$ &  $ \tilde{\delta} d_e(sys)$ & $ \delta d_e(sys)$  \\
\hline
   20. & -1996.   & -213498.  &  -320.   & -1663.& -78.    &  -33.   \\ 
   30. & -658.    &  -42436.  &  -140.   &  -495.& -35.    &  -9.9    \\
   40. & -342.    &  -15657.  &  -79.    & -211. & -20.    &  -4.2    \\      
   50. & -212.    &   -7546.  &  -50.    & -107. & -13.    &  -2.1    \\
   60. & -144.    &   -4229.  &  -35.    & -61.  &  -8.7   &  -1.2    \\
   70. & -105.    &   -2613.  &  -26.    & -39.  &  -6.4   &  -0.78   \\
   80. &  -80.    &   -1728.  &  -20.    & -26.  &  -4.9   &  -0.52   \\
   90. &  -63.    &   -1204.  &  -15.    & -18.  &  -3.9   &  -0.37   \\
  100. &  -51.    &    -872.  &  -13.    & -14.  &  -3.1   &  -0.27   \\
  110. &  -42.    &    -653.  &  -10.    & -10.  &  -2.6   &  -0.20   \\
  120. &  -35.    &    -501.  &   -8.7   & -7.7  &  -2.2   &  -0.15   \\
  130. &  -30.    &    -393.  &   -7.4   & -6.1  &  -1.9   &  -0.12   \\
  140. &  -26.    &    -314.  &   -6.4   & -4.9  &  -1.6   &  -0.098  \\
  150. &  -22.    &    -255.  &   -5.6   & -4.0  &  -1.4   &  -0.080  \\
  160. &  -20.    &    -210.  &   -4.9   & -3.1  &  -1.2   &  -0.063  \\
  170. &  -17.    &    -175.  &   -4.3   & -2.7  &  -1.1   &  -0.055  \\
  180. &  -15.    &    -147.  &   -3.9   & -2.2  &   -0.97 &  -0.044  \\ 
  190. &  -14.    &    -125.  &   -3.5   & -1.9  &   -0.87 &  -0.038  \\
  200. &  -13.    &    -107.  &   -3.1   & -1.6  &   -0.78 &  -0.032  \\
\hline
\end{tabular}
\label{dde2}
\end{table}

%%%%%%%%%%%%%%%%%%%%%%%%%%%%%%%%%%%%%%%%%%%%%%%%%%%%%%%%%%%%%%%%%%%%%%%%%%%%%%%
%%%%%%%%%%%%%%%%%%%%%%%%%%%%%%%%%%%%%%%%%%%%%%%%%%%%%%%%%%%%%%%%%%%%%%%%%%%%%%%

The author is grateful to A. V. Titov for useful discussions.
Codes for calculations of diatomic molecules were developed with the support of the Russian Science Foundation grant (project
No. 14-31-00022).
Calculation of the ThO molecule were performed with the
support of 
Saint Petersburg State University,
research grant 0.38.652.2013 and RFBR Grant No. 13-02-0140.
\section{Appendix}
\label{app}

In the first order in the small parameter $\Delta E_{ST}/ \Delta E_{rot} \sim \E d/\Delta E_{rot}$ for $\delta^{1} \Delta E_{ST}^{u(l)}$,  $\delta^{2} \Delta E_{ST}^{u(l)}$, $\delta^{1}B^{u(l)}$, and  $\delta^{2}B^{u(l)}$
we have
\begin{eqnarray}
\nonumber 
  \delta^{1} \Delta E_{ST}^{u(l)} = \\
\nonumber
 \E^2 /\Delta E_{\rm rot} |<J{=}1,{M{=}1,\Omega}|d_z| J'{=}2,{M=1},\Omega>|^2 = \\
\nonumber
 \frac{\E^2 d^2}{\Delta E_{\rm rot}} \frac{((J{+}1)^2-M^2)((J{+}1)^2-\Omega^2)} { (2J{+}1)(2J{+}3)(J{+}1)^2},
\end{eqnarray}

\begin{eqnarray}
\nonumber 
  \delta^{2} \Delta E_{ST}^{u(l)} = \\
\nonumber
 -\E^2 /\Delta E_{\rm rot} |<J{=}1,{M{=}0,\Omega}|d_z| J'{=}2,{M=0},\Omega>|^2 = \\
\nonumber
-\frac{\E^2 d^2}{\Delta E_{\rm rot}} \frac{(J{+}1)^2((J{+}1)^2-\Omega^2)} { (2J{+}1)(2J{+}3)(J{+}1)^2},
\end{eqnarray}

\begin{eqnarray}
\nonumber 
\delta^{1}B_{u(l)} = \E_{\perp}/2 <J'{=}2,{M{=}1,\Omega}|d_+| J{=}1,{M'{=}0}, \Omega> \times \\
\nonumber
\E /\Delta E_{\rm rot} <J{=}1,{M=1,\Omega}|d_z| J'{=}2,{M=1},\Omega>  = \\
\nonumber
- \frac {\E_{\perp} \E d^2}{2\Delta E_{\rm rot}}
 \sqrt{ \frac {(J{+}M)(J{+}M{+}1) ((J{+}1)^2-\Omega^2)}{(2J{+}1) (2J{+}3)(J{+}1)^2}} \times \\
\nonumber
 \sqrt{\frac{((J{+}1)^2-M^2)((J{+}1)^2-\Omega^2)}{(2J{+}1)(2J{+}3)(J{+}1)^2}}, 
\end{eqnarray}

\begin{eqnarray}
\nonumber 
\delta^{2}B_{u(l)} = \E_{\perp}/2 <J{=}1,{M{=}1,\Omega}|d_+| J'{=}2,{M'{=}0}, \Omega> \times \\
\nonumber
 \E / \Delta E_{\rm rot} <J{=}1,{M'{=}0,\Omega}|d_z| J'{=}2,M'{=}0>  = \\
\nonumber
 \frac {\E_{\perp} \E d^2}{2 \Delta E_{\rm rot}}
 \sqrt{ \frac {(J{-}M{+}2)(J{-}M{+}1)((J{+}1)^2-\Omega^2)}{(2J{+}1) (2J{+}3)(J{+}1)^2}} \times \\
\nonumber
 \sqrt{\frac{(J{+}1)^2((J{+}1)^2-\Omega^2)}{(2J{+}1)(2J{+}3)(J{+}1)^2}}. 
\end{eqnarray}
$\Delta E_{\rm rot}$ is negative, therefore $\delta^{(1)} \Delta E_{ST}^{u(l)} {<} 0$ and $\delta^{(2)} \Delta E_{ST}^{u(l)} {>} 0$.

Then retaining terms up to the first order in $\Delta E_{ST}/ \Delta E_{rot}$ we have

\begin{eqnarray}
\label{cancel2}
\nonumber
\frac {B_{u(l)}^2} {A_{u(l)}^2}  = \frac{(B + \delta^{1}B_{u(l)} + \delta^{2}B_{u(l)})^2} {(\Delta E_{ST} + \delta^{1} \Delta E_{ST}^{u(l)} + \delta^{2} \Delta E_{ST}^{u(l)})^2} \approx \\
\nonumber
\frac{B^2 + 2B\delta^{1}B_{u(l)} + 2B\delta^{2}B_{u(l)}} {\Delta E_{ST}^2 + 2\Delta E_{ST}\delta^{1} \Delta E_{ST}^{u(l)} + 2\Delta E_{ST}\delta^{2} \Delta E_{ST}^{u(l)}} = \\
\nonumber
\frac{\frac {\E_{\perp}^2 d^2 \Omega^2} {4} \frac {(J{-}M{+}1)(J{+}M)} {(J(J{+}1))^2} }
{\frac{\E^2 d^2 M^2\Omega^2}{(J(J{+}1))^2}   } \times\\
\nonumber
\frac{  \left( 1  + \left[(J{+}M{+}1) - (J{+}1)\sqrt{\frac{J{-}M{+}2}{J{+}M}}\right]K(J)/\Omega\right)  }
{ \left( 1 +MK(J)/\Omega       \right)  } = \\
\frac{B^2  \left( 1  {+} \left[(J{+}M{+}1) {-} (J{+}1)\sqrt{\frac{J{-}M{+}2}{J{+}M}}\right]K(J)/\Omega\right)  }
{A^2 \left( 1 +MK(J)/\Omega       \right)  },
\end{eqnarray}
where
\begin{equation}
\nonumber
K(J)=  2\frac{\E d}{\Delta E_{\rm rot}} \frac{J(J{+}1)((J{+}1)^2-\Omega^2)}{(2J{+}1)(2J{+}3)(J{+}1)^2}.
\end{equation}
Substituting $M{=}1$, $J{=}1$ to Eq. (\ref{cancel2}) we have got Eq. (\ref{cancel}).
Eq. (\ref{cancel2}) is obtained for $M{=}{+}1$ level. The result is the same for $M{=}{-}1$.
Taking into account that $|J{=}1,M{=}+1,\Omega{=}+1>$ ($|J{=}1,M{=}+1,\Omega{=}-1>$) corresponds to the upper (lower) $\Omega$-doublet level,  for the next term we have
$$\frac { B_{u}^2} { A_{u}^4} = \frac { B^2(1+K)} { A^4(1+2K)} = \frac { B^2(1+\frac{\E d}{5\Delta E_{\rm rot}})} { A^4(1+2\frac{\E d}{5\Delta E_{\rm rot}} )}, $$
$$\frac { B_{l}^2} { A_{l}^4} = \frac { B^2(1-K)} { A^4(1-2K)} = \frac { B^2(1-\frac{\E d}{5\Delta E_{\rm rot}})} { A^4(1-2\frac{\E d}{5\Delta E_{\rm rot}} )}. $$

As an example, below are the parameters calculated  for $\E = 110$ V/cm.
$$\frac {2\hbar B_{u}^2} {4\Eeff A_{u}^2}  =  10.33057851 \times 10^{-29}\frac{\ecm}{2\pi\rm MHz}, $$
$$\frac {2\hbar B_{l}^2} {4\Eeff A_{l}^2}  =  10.33057851 \times 10^{-29}\frac{\ecm}{2\pi\rm MHz}, $$

$$\frac {2\hbar^3 B_{u}^2} {4\Eeff A_{u}^4} =   0.00080156 \times 10^{-29}\frac{\ecm}{(2\pi\rm MHz)^3}, $$
$$\frac {2\hbar^3 B_{l}^2} {4\Eeff A_{l}^4}  =   0.00080343 \times 10^{-29}\frac{\ecm}{(2\pi \rm MHz)^3}. $$

%\bibliographystyle{./bib/apsrev}
%\bibliography{bib/JournAbbr,bib/SkripnikovLib,bib/QCPNPI,bib/TitovLib,bib/Kaldor,bib/PetrovLib,bib/Lomachuk,bib/ACME}

\end{document}